\documentclass[twocolumn,fp]{jpsj3}
\usepackage{txfonts}

\title{
Structural quantum criticality and superconductivity\\
in iron-based superconductor Ba(Fe$_{1-x}$Co$_x$)$_2$As$_2$
}

\author{\name{Masahito \surname{Yoshizawa}}\thanks{E-mail address:yosizawat@iwate-u.ac.jp}$^{1}$, 
\name{Daichi \surname{Kimura}}$^{1,4}$, 
\name{Taiji \surname{Chiba}}$^{1}$, 
\name{Abdusalam \surname{Ismayil}}$^{1}$, \\
\name{Yoshiki \surname{Nakanishi}}$^{1,4}$, 
\name{Kunihiro \surname{Kihou}}$^{2,4}$, 
\name{Chul-Ho \surname{Lee}}$^{2,4}$, 
\name{Akira \surname{Iyo}}$^{2,4}$, \\
\name{Hiroshi \surname{Eisaki}}$^{2,4}$, 
\name{Masamichi \surname{Nakajima}}$^{3,4}$, and 
\name{Shin-ichi \surname{Uchida}}$^{3,4}$, 
}

\inst{$^{1}$ Graduate School of Engineering, Iwate University, Morioka 020-8551, Japan\\
$^{2}$ National Institute of Advanced Industrial Science and Technology (AIST), Tsukuba 305-8568, Japan \\
$^{3}$ Graduate School of Science, The University of Tokyo, Tokyo 113-0033, Japan \\
$^{3}$ Transformative Research-project on Iron Pnictides (TRIP), Japan Science and Technology Agency, Tokyo 102-0075, Japan
}

\abst{
We investigated the elastic properties of the iron-based superconductor Ba(Fe$_{1-x}$Co$_x$)$_2$As$_2$ with eight Co concentrations. 
The elastic constant $C_{66}$ shows large elastic softening associated with the structural phase transition.
The $C_{66}$ was analyzed base on localized and itinerant pictures of Fe-3d electrons, which shows the strong electron-lattice coupling and a possible mass enhancement in this system.
The results resemble those of unconventional superconductors, where the properties of the system are governed by the quantum fluctuations associated with the zero-temperature critical point of the long-range order; namely, the quantum critical point (QCP). 
In this system, the inverse of $C_{66}$ behaves just like the magnetic susceptibility in the magnetic QCP systems.
While the QCPs of these existing superconductors are all ascribed to antiferromagnetism, our systematic studies on the canonical iron-based superconductor Ba(Fe$_{1-x}$Co$_x$)$_2$As$_2$ have revealed that there is a signature of ''structural quantum criticality'' in this material, which is so far without precedent. 
The elastic constant anomaly is suggested to concern with the emergence of superconductivity.
These results highlight the strong electron-lattice coupling and effect of the band in this system, thus challenging the prevailing scenarios that focus on the role of the iron 3d-orbitals.
}

\kword{iron-based superconductor, elastic constant, quantum criticality, electron-lattice coupling}

\begin{document}
\maketitle

\section{Introduction} 
\begin{figure}[b]
\begin{center}
\includegraphics[width=8.5cm]{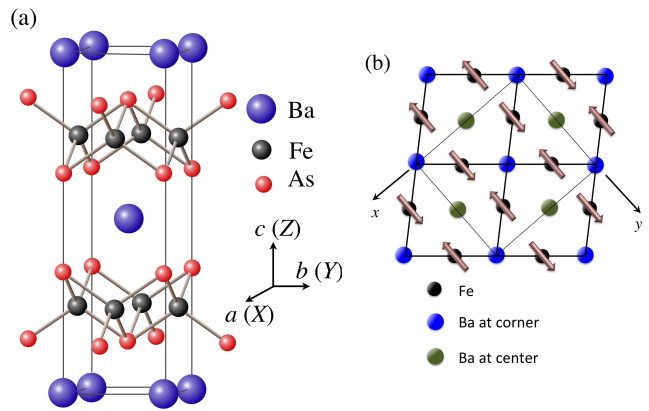}
\end{center}
\caption{(Color online) (a) Crystal structure of BaFe$_2$As$_2$, which belongs to the base-centered tetragonal crystal class $I_{4/mmm}$, and (b) crystal structure and magnetic order below $T_{\rm S}$ with the crystal symmetry of $F_{mmm}$. }
\label{crystal_structure}
\end{figure}

Iron-based superconductors have been studied actively worldwide as new high-temperature superconductors. 
Their superconducting transition temperature is as high as that of cuprates, and superconductivity emerges in the system with the ferromagnetic Fe element \cite{jonston}. 
These characteristics are evoking global interest in the mechanism of superconductivity \cite{jonston}. 
Among the many iron-based superconductors, Ba(Fe$_{1-x}$Co$_x$)$_2$As$_2$ is suitable for basic research, because it provides high-quality large single crystals. 
The crystal structure of the parent compound, BaFe$_2$As$_2$, is shown in Fig. \ref{crystal_structure}(a) \cite{rotter}. 
It has a tetragonal crystal structure at room temperature. 
This becomes an orthorhombic structure at the structural transition temperature $T_{\rm S}$ = 140 K accompanying the appearance of the long-range antiferromagnetic order \cite{rotter} as shown in Fig. \ref{crystal_structure}(b), which is viewed from the c-axis \cite{huang}. 
$T_{\rm S}$ decreases when Co is substituted for Fe, and superconductivity appears at $x$ = 0.03 \cite{sefat,rotter2}. 
$T_{\rm S}$ falls to zero at $x$ = 0.07, where the superconducting temperature $T_{\rm sc}$ becomes highest, reaching 25 K. 
The magnetic ordering temperature $T_{\rm N}$ coincides with $T_{\rm S}$ for $x$ = 0, and for samples in which $x$ is less than 0.03, $T_{\rm N}$ is always lower than $T_{\rm S}$. 
Such a phase diagram showing the coexistence of magnetic order and superconductivity has been observed in the rare-earth compound CePd$_2$Si$_2$ and the uranium compound UGe$_2$, where superconductivity appears near the quantum critical point (QCP) \cite{mathur,saxena}. 

Intensive studies on QCP have led to a new paradigm in physics in which magnetic fluctuation mediates superconductivity. 
In the case of iron-based superconductors, the mechanism of superconductivity has so far been discussed based on the origin of the adjacent magnetic and structural phases, whether magnetic or orbital \cite{fang,krueger}. 
Mazin {\it et al}. and Kuroki {\it et al}. predicted that a fully gapped sign-reversing s-wave state ($s_ \pm$ state) would be realized by magnetic instability \cite{mazin,kuroki}. 
On the other hand, Kontani and Onari, and Yanagi {\it et al}. proposed a conventional s-wave state without sign reversal ($s_{++}$ state) mediated by orbital fluctuation \cite{kontani1,yanagi1}. 
Investigations on the neighboring order and its fluctuation are important to clarify the superconducting properties of iron-based superconductors.

Our experimental tool, ultrasonic measurements, provides us with information on changes in the symmetry of the lattice system. 
Strains introduced into the crystal in the elastic constant measurements deform the crystal locally and break the crystal symmetry (symmetry breaking field). 
When the system encounters either a charge, or orbital (electric quadrupole) order, the elastic stiffness that possesses the same symmetry as, and thus couples with, this order tends to exhibit anomaly (softening in most cases). 
Accordingly, by examining the temperature ($T$)-dependence of the anisotropic elastic stiffness $C_{ij}$, we can obtain fundamental information on the symmetry of the order. 

The T-dependence of the elastic stiffness of Ba(Fe$_{1-x}$Co$_x$)$_2$As$_2$ has been reported for platy samples of BaFe$_2$As$_2$ and Ba(Fe$_{0.8}$Co$_{0.2}$)$_2$As$_2$ using the resonant ultrasonic spectroscopy (RUS) method \cite{fernandes} and for a bulk sample of overdoped Ba(Fe$_{0.9}$Co$_{0.1}$)$_2$As$_2$ using pulsed ultrasonic spectroscopy with a phase comparison method \cite{goto}. 
The former study reported a large elastic softening toward $T_{\rm S}$ in $C_{66}$, which was present at the percentage ratios of 95\% and 18\% in BaFe$_2$As$_2$ and Ba(Fe$_{0.8}$Co$_{0.2}$)$_2$As$_2$, respectively. 
The latter study reported that $C_{11}$, $C_{33}$, $C_{44}$, and $\frac{1}{2}\left( {C_{11}  - C_{12} } \right)$ gradually increased as $T$ decreased, and showed no remarkable anomaly at $T_{\rm sc}$ in contrast with the $C_{66}$ anomaly of 21\%. 

In the present study, we focused our attention on the huge anomaly in $C_{66}$ and measured the $T$-dependence of Ba(Fe$_{1-x}$Co$_x$)$_2$As$_2$ with eight Co concentrations from $x$ = 0 to $x$ = 0.245 to investigate the neighboring phase and its role in the superconducting properties. 

\section{Experimental}
\label{exp}
\subsection{Ultrasonic measurements}
We measured the elastic constants for Ba(Fe$_{1-x}$Co$_x$)$_2$As$_2$ using an ultrasonic pulse-echo phase comparison method \cite{luethi1} as a function of temperature from 5 K to 300 K using a cryostat mounted on a Gifford-McMahon (GM) cryocooler. 
Ultrasound was emitted and detected by LiNbO$_3$ transducers. 
X-cut plate of LiNbO$_3$ at 41$^\circ$ with a thickness of 100 $\mu$m was used for the transverse waves. 
The fundamental resonance frequency was 19 MHz. 
We usually use their higher harmonics of 67 MHz for the experiments. 
The transducers were glued onto a pair of parallel planes of single crystals using an elastic polymer Thiokol. 

Elastic stiffness was obtained by $C$ = $\rho v^2$, where $\rho$ is the density and $v$ is either the longitudinal or transverse sound velocity.
The corresponding sound velocities can be obtained by choosing the propagation and displacement directions.
Tetragonal crystal symmetry has six independent $C_{ij}$'s; namely, $C_{11}$, $C_{33}$, $C_{12}$, $C_{13}$, $C_{44}$, and $C_{66}$.
The propagation and displacement directions of the sound velocity was [100] and [010] for $C_{66}$.
The sound velocity was obtained by the time interval of the echo train and the sample length, whose accuracy is within a few percent due to the usage of large crystals. 
The value of $\rho$ was obtained using the data of $x$ = 0 and 0.1 \cite{sefat}. 
By assuming Vegard$^\prime$s law, the lattice constant of the $a(b)$ and $c$ axes are $a$ = $b$ = 0.39636 + 3.8981 ~ 10$^{-4}$ $x$ (nm) and $c$ = 1.3022 - 0.0421$x$ (nm), respectively, for Ba(Fe$_{1-x}$Co$_x$)$_2$As$_2$. 
The value of $\rho$ is 6.48 x 10$^3$ kg$\cdot$m$^{-3}$ and 6.55 ~ 10$^3$ kg$\cdot$m$^{-3}$ for $x$ = 0 and 0.245, respectively. 
The $XYZ$ coordinate was defined by the unit cell of the $I_{4/mmm}$ crystal structure \cite{rotter}. 
To prevent damage to the sample due to rapid changes in temperature, the rate of change in temperature was carefully controlled so as to be 10 K/h near $T_{\rm S}$ \cite{huang}.

High-quality large single crystals of Ba(Fe$_{1-x}$Co$_x$)$_2$As$_2$ used in this work were grown by the self-flux method. 
Samples with eight Co concentrations $x$ = 0, 0.037, 0.060, 0.084, 0.098, 0.116, 0.161, and 0.245 were prepared. 
The Co concentration in the grown crystals was determined by energy-dispersive X-ray spectroscopy (EDS) measurement. 
The Co content of the samples was about 75\% of the prepared one. 
The samples were cut into a rectangular shape, after determining their axis by X-ray Laue photography. 
The typical size of a sample was 3 mm $\times$ 3 mm in the tetragonal $ab$ ($XY$) cleavage plane, and 2 mm in thickness on the $c(Z)$-axis.

\subsection{Temperature dependence of $C_{66}$}
\begin{figure}[t]
\begin{center}
\includegraphics[width=8.5cm]{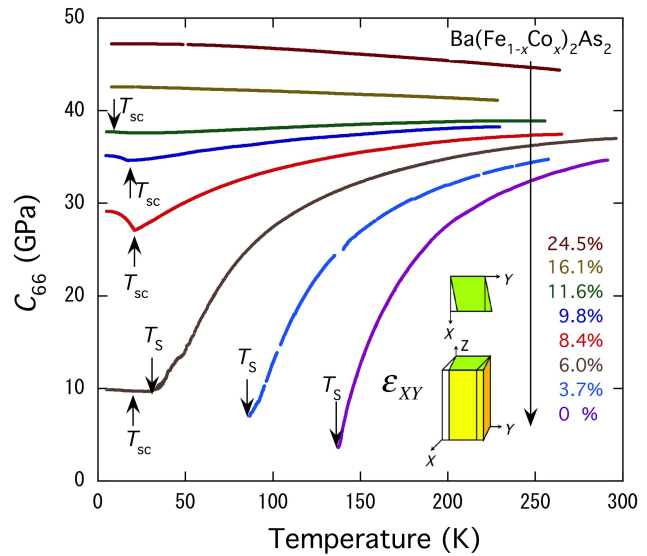}
\end{center}
\caption{(Color online) Temperature-dependence of the elastic stiffness $C_{66}$ of Ba(Fe$_{1-x}$Co$_x$)$_2$As$_2$ with various values.}
\label{C66}\end{figure}

Figure \ref{C66} shows the $T$-dependence of $C_{66}$ for samples with $x$ = 0, 0.037, 0.060, 0.084, 0.098, 0.116, 0.161, and 0.245. 
$C_{66}$ significantly decreases as $T$ decreases. 
While a decrease in elastic stiffness is a common feature as a precursor of the structural phase transition, the amount of softening, as much as 80\% for Ba(Fe$_{0.963}$Co$_{0.037}$)$_2$As$_2$, is unprecedentedly large. 
The softening in $C_{66}$ corresponds to the symmetry change from tetragonal to orthorhombic. 
This is consistent with the structural analysis of this material, where the space group is $I_{4/mmm}$ and $F_{mmm}$ for the high- and low-temperature phases, respectively \cite{rotter}.

The decrease in $C_{66}$ with decreasing $T$ is prominent for $x$ = 0, 0.037, and 0.060, which follow the disappearance of the signal at the structural transition temperature $T_{\rm S}$ = 141 K ($x$ = 0), 84.7 K ($x$ = 0.037), and 30 K ($x$ = 0.060). 
The data below $T_{\rm S}$ were not plotted for $x$ = 0 and 0.037, because the sound echo signal disappeared at a certain temperature range below $T_{\rm S}$, which may be ascribed to strong scattering of the sound by the orthorhombic domain boundaries.
The softening of $C_{66}$ is less prominent as $x$ increases, which is consistent with the disappearance of the structural phase transition. 
For the superconducting samples, 0.060 $\leqslant$ $x$ $\leqslant$ 0.116, anomalies are observed at $T_{\rm sc}$ = 24.0 K ($x$ = 0.060), 20.7 K ($x$ = 0.084), 16.5 K ($x$ = 0.098), and 10.5 K ($x$ = 0.116). 

The underdoped samples show a peak at $T_{\rm sc}$ and a step-like anomaly below $T_{\rm sc}$, as shown in Fig. \ref{C66large}. 
On the other hand, $C_{66}$ increases below $T_{\rm sc}$ in the overdoped region. 
This difference will be discussed later. 
Moreover, the superconducting transition disappears when $x$ $>$ 0.161. 
$C_{66}$ for the samples of $x$ = 0.161 and 0.245 shows gradual increases as $T$ decreases, which a typical behavior for any material, reflecting phonon anharmonicity.

\section{Discussion}
\label{analysis}
\subsection{Analysis based on localized picture}

\begin{figure}[t]
\begin{center}
\includegraphics[width=8.5cm]{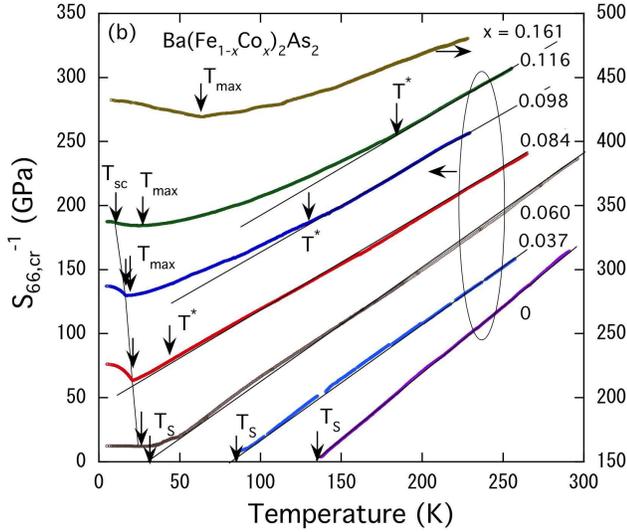}
\end{center}
\caption{(Color online) Temperature-dependence of the inverse of $S_{66,cr}$.}
\label{inv_S66}
\end{figure}

The observed large elastic anomaly in $C_{66}$ is a precursor of a structural phase transition. 
Here, we will discuss the origin of the softening in $C_{66}$ from the viewpoint of experimentalists to provide the analysis method \cite{yoshizawa_news}.
The 3d orbitals are split into $E_g$ doublet and $T_{2g}$ triplet by the electric crystalline field (CEF) in a cubic symmetry. 
The $E_g$ is spilt into two singlets, and the $T_{2g}$ is split into one singlet and one doublet by tetragonal CEF of iron-based materials. 
The remaining doublet can be lifted by the elastic strain, and may cause the elastic anomaly. 

We will consider the coupling between the strain $\varepsilon$ and an order parameter $O$
as $H =  - \lambda O_\Gamma  \varepsilon _\Gamma $.
The equivalence of the $X$ and $Y$ axes in tetragonal symmetry leads to the degeneration of the d$_{ZX}$ and d$_{YZ}$ orbitals. 
This degeneracy is lifted by $\varepsilon_{XY}$ or $\varepsilon _{XX} - \varepsilon _{YY}$, and brings about anomalies in the corresponding $C_{66}$ and $\frac{1}{2}\left( {C_{11}  - C_{12} } \right)$. 
The strain $\varepsilon_{XY}$ can couple with the orbital (quadrupole) $O_{XY}$. 
We can calculate the strain susceptibility based on the localized picture of d electron.

\begin{equation}
\label{C}
C_{66}   = C_{66,0}  - N\lambda ^2 \frac{{\chi _{66} ^0 }}
{{1 - I\chi _{66}^0 }} = C_{66,0} \frac{{T - T_{\rm{C}} }}
{{T - \Theta }}
\end{equation}

\noindent where $\lambda$,  $I$, and $N$ are the coupling constant, the intersite interaction, and the number of atoms per unit volume. 
Here, we adopted the form of $\chi _{66}^0  = \frac{{\left\langle {\left( {O_{XY}} \right)^2 } \right\rangle }}{T}$, for localized d electrons.

Here, we introduce the elastic compliance $S_{ij}$, which is a component of the inverse $C_{ij}$ matrix. 
In Eq. (\ref{C}), the transition undergoes at $T_{\rm C}$, where the lattice shows an instability. 
Elastic compliance represents the gstructuralh susceptibility of elastic systems, and corresponds to the magnetic susceptibility $\chi$ in magnetic systems. 
The experimentally observed $S_{66}$ (= 1/$C_{66}$) can be decomposed into the sum of the anomalous contributions that exhibit critical behavior $S_{66,cr}$ and the normal contribution (background) $S_{66,0}$. 

\begin{equation}
\label{JT}
S_{66} = S_{66}^0 + S_{66,cr} = S_{66,0} \left( {1 + \frac{{E_{{\rm{JT}}} }}{{T - T_{\rm{C}} }}} \right)
\end{equation}

\noindent where $E_{\rm JT} = T_{\rm{C}}  - \Theta$. 
$E_{\rm JT}$ stands for the Jahn-Teller energy, an energy scale that corresponds to the strength of the electron-lattice coupling. 
Note that this formula has the same form as the Curie-Weiss susceptibility of ferromagnetic materials. 

For the analysis, we employed the data on Ba(Fe$_{0.755}$Co$_{0.245}$)$_2$As$_2$ for $S_{66,0}$, and subtract it from the other data. 
Figure \ref{inv_S66} shows the inverse of $S_{66,cr}$ as a function of temperature. 
It can clearly be seen that 1/$S_{66,cr}$ in the underdoped region ($x$ $<$ 0.070) exhibits linear $T$-dependence. 
This indicates that $S_{66,cr}$ obeys Eq. (\ref{JT}). 

In the underdoped samples, $T_{\rm C}$ in Eq. (\ref{JT}) and $T_{\rm S}$ obtained by $C_{66}$ are 141 K and 136 K for $x$ = 0, 84 K and 81.4 K for $x$ = 0.037, and 30 K and 28.2 K for $x$ = 0.060, respectively. 
The closeness of the values in each case suggests the occurrence of ferro-order. 
This strongly suggests the crystal deformation of $\varepsilon_{XY}$ below $T_{\rm S}$.

For the underdoped samples, $E_{\rm JT}$ is approximately equal to 50 K.
This value is comparable to $\Delta$ = 20 K in Ref. [15], which is smaller than our value.
The difference between the two experiments is originated from how to choose the background C$_0$. 
While we analyzed the data without a fitting parameter for the background by adopting the data of Ba(Fe$_{0.755}$Co$_{0.245}$)$_2$As$_2$ for $C_0$, 
Goto {\it et al}. carried out the fitting including the background as adjustable parameters. 
We performed the analysis by using the same $C_0$, which can reveal the systematic behavior though the whole Co concentration.
The values of the electron-lattice coupling constant $\lambda \sqrt {\left\langle {\left( {O_{XY} } \right)^2 } \right\rangle }$ was calculated from $E_{\rm JT}$, and is 0.25 eV/Fe for $x$ = 0.037.

\subsection{Analysis based on band picture}
\begin{figure}[t]
\begin{center}
\includegraphics[width=8.5cm]{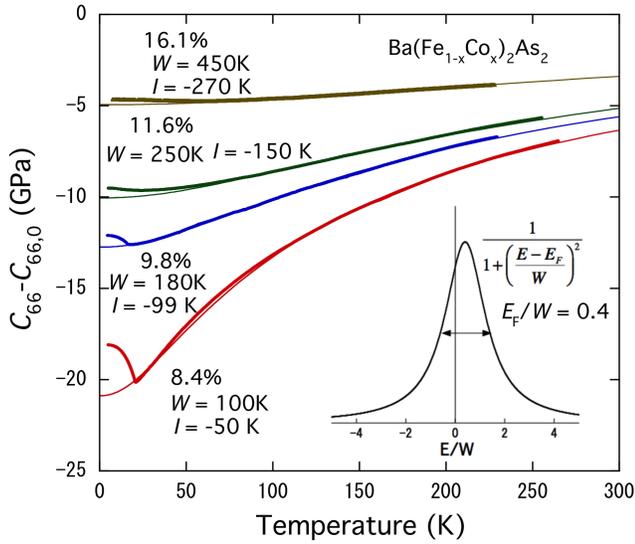}
\end{center}
\caption{(Color online) Theoretical fitting of $C_{66}$ by the band Jahn-Teller effect under the assumption of Lorentzian density of states. 
Adjustable parameters are the bandwidth $W$ and the intersite interaction $I$.}
\label{bandanalysis}
\end{figure}

When $x$ exceeds 0.07, $S_{66,cr}$ deviates from $T$-linear behavior below a certain temperature $T^*$; i.e., 40 K for $x$ = 0.084, 130 K for $x$ = 0.098, and 190 K for $x$ = 0.116. 
The $T$-dependence of $C_{66}$ in the overdoped region can be explained by a band picture instead of Eq. (\ref{JT}). 
Large elastic anomalies compared with iron-based materials have been reported in the A15 superconductor V$_3$Si and the Laves-phase superconductor CeRu$_2$ so far \cite{testerdi,suzuki}. 
These anomalies have been ascribed to the large density of states at the Fermi energy. 
The 3d-orbitals form bands in an iron-based superconductor. 
The bands located above the Fermi energy at the ƒ¡-point form hole Fermi surfaces and electron pockets at M-points of the zone boundary \cite{yanagi1}. 
Four M-points (which are identical to X-points in the square lattice formed by Fe atoms) do not becomes equivalent under $\varepsilon_{XY}$ \cite{yoshizawa_news}. 
If a large density of states exists at the M-point, an elastic anomaly may be caused. 
This provides a natural solution to the question of why the anomaly only emerged in $C_{66}$ \cite{yoshizawa_news}. 
The formula for the elastic constant based on this consideration is as follows \cite{thalmeier,yoshizawa_bp}:

\begin{equation}
\label{band_JT}
C = C_0  - \left( {d_{M1}  - d_{M2} } \right)^2 \frac{{\chi _S^0 }}{{1 - I\chi _S^0 }}
\end{equation}

\[
\begin{array}{lll}
\chi _S^0  = \frac{1}
{{k_{\rm{B}} T}}\sum\limits_k {f_k \left( {1 - f_k } \right)}  = \int {dE\,N\left( E \right)} \,f\left( E \right)\\
N\left( E \right) = \frac{{N_0 }}
{2}\left[ {1 + \left( {\frac{{E - E_{\rm{F}} }}
{W}} \right)^2 } \right]^{ - 1}\\ 
\end{array} 
\]
	
\noindent where $f$ is the Fermi-Dirac function, $d$ (= $d_{M1}$ = -$d_{M2}$) is the electron-lattice coupling constant, $N_0$ is the density of states at Fermi energy $E_{\rm F}$, and $I$ is the intersite interaction. 
The results of fitting by adopting the Lorentzian density of states are shown in Fig. \ref{bandanalysis}. 

In this analysis, the main adjustable parameter is the bandwidth $W$. 
We fixed $E_{\rm F}$/$W$ = 0.4 to obtain the best fitting. 
The value of $I$ is displayed in the figure, depending on the value of $x$. 
We added a small constant ranging from 0.8 GPa to 2.45 GPa to the experimental value of $C_0$ for the fitting. 
The value of $W$ was 100, 180, 250, and 450 K for the 8.4\%, 9.8\%, 11.6\%, and 16.1\% samples, respectively. 
From the adjustable parameter $N_0d^2$, the value of $d$ was respectively evaluated as 0.22, 0.25, 0.28, and 0.28 eV/Fe for the four samples, under the assumption that $N_0$ is approximately equal to $1/W$.
The values of $N_0$ calculated from $N_0$ = 1/$W$ are respectively 110, 64, 46, and 26 states/eV for the four samples. 

\begin{figure}[t]
\begin{center}
\includegraphics[width=8.5cm]{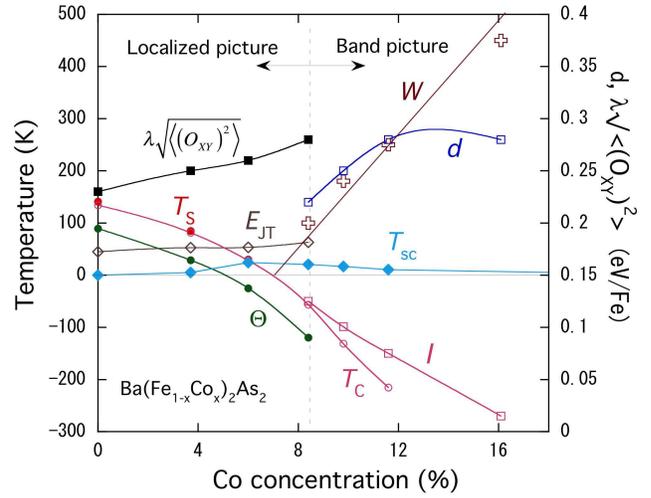}
\end{center}
\caption{(Color online) The obtained parameters from the analyses based on the localized and band pictures of Eqs. \ref{JT} and \ref{band_JT}, respectively. 
Electron-lattice coupling constants of $\lambda \sqrt {\left\langle {\left( {O_{XY} } \right)^2 } \right\rangle }$  was evaluated from $E_{\rm JT}$. 
Evaluation method of $d$ is described in the text. 
The curves are the guides for eyes.}
\label{parameterfig}
\end{figure}

The band effect discussed here is called the electronic redistribution mechanism \cite{luethi159}, and is applicable to systems having low resistivity. 
It is supported experimentally. The resistivity of this system becomes lower in the overdoped region compared with the underdoped region \cite{rullier,nakajima}. 
Band nesting along the $[\pi/a, \pi/a, 0]$ direction is a characteristic feature of iron-based superconductors. 
We have no tool for analyzing elastic data based on band nesting. 
However, we infer that it has a similar effect to that discussed above, and may cause $C_{66}$ softening. 
The obtained parameters are summarized in Fig. \ref{parameterfig}

\subsection{Quantum criticality}
\subsubsection{Phase diagram}
\begin{figure}[t]
\begin{center}
\includegraphics[width=8.5cm]{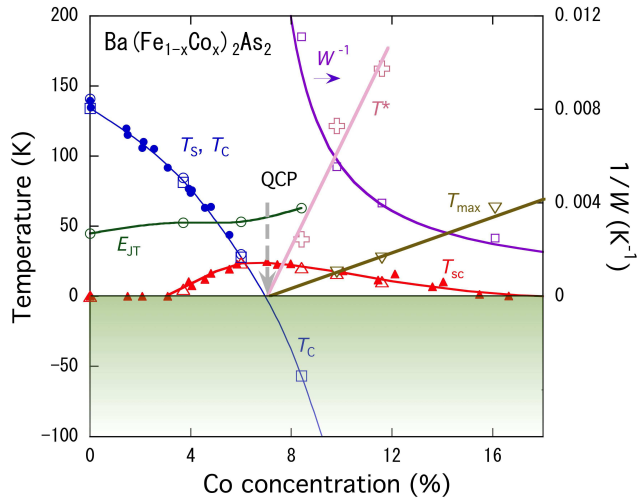}
\end{center}
\caption{(Color online) Phase diagram of Ba(Fe$_{1-x}$Co$_x$)$_2$As$_2$. 
Filled circles: $T_{\rm S}$, and filled triangles: $T_{\rm sc}$ from other studies \cite{canfield,laplace}. 
Open symbols in this study. 
The curves are the guides for eyes.}
\label{phasediagram}
\end{figure}

Figure \ref{phasediagram} summarizes the phase diagram of the Ba(Fe$_{1-x}$Co$_x$)$_2$As$_2$. 
The Co concentration-dependence of $T_{\rm S}$ and $T_{\rm sc}$ are highly consistent with the values reported elsewhere \cite{canfield,laplace}. 
We found two characteristic temperatures of $T^*$ and $T_{\rm max}$. 
A possible explanation of $T^*$ is the crossover from the non-Fermi liquid to the Fermi liquid region. 
The boundary from non-Fermi $T$-dependence to Fermi liquid $T^2$-dependence observed in the resistivity measurements is not clear for Ba(Fe$_{1-x}$Co$_x$)$_2$As$_2$ \cite{doiron}. 
However, the behavior  observed in  BaFe$_2$(As$_{1-x}$P$_x$)$_2$ is similar to the case in the present study including the $x$-dependence of the crossover region  \cite{kasahara}. 

We found $T_{\rm max}$, which corresponds to the temperature at which $S_{66,cr}$ takes the maximum value (1/$S_{66,cr}$ takes the minimum value), as shown in Figs. \ref{inv_S66} and \ref{S66}. 
For highly correlated electron systems such as CeCu$_2$Si$_2$, UPd$_2$Al$_3$, and UPt$_3$, a similar maximum was reported in the magnetic susceptibility $\chi$,  and is interpreted as a Kondo temperature, signaling the coherent motion of f-electrons \cite{onuki}. 
From this analogy, $T_{\rm max}$ in this system may suggest the onset of new coherent states. 

It would be notable that  $T^*$ and $T_{\rm max}$ go to zero at the QCP concentration of $x_{\rm C}$ = 0.07, indicating the existence of structural QCP at this concentration. 
In addition, the analysis based on the band picture suggests that the bandwidth $W$ also goes to zero at the same $x_{\rm C}$, which suggest a possible mass enhancement toward QCP in this system. 
Note that this point is located at the center of the superconducting dome. 
A similar phenomenon was reported for BaFe$_2$(As$_{1-x}$P$_x$)$_2$ \cite{nakai}. 
The obtained phase diagram and various phenomena near the QCP resemble those of the well-known rare-earth compounds and uranium compounds. 
This coincidence strongly suggests the intimate relationship between superconductivity and QCP in this system, similarly to heavy fermion systems where magnetic QCP is believed to be responsible for the emergence of unconventional superconductivity \cite{mathur}. 
The essential difference in this case is that the quantum criticality is associated with structural fluctuations, unlike the magnetic fluctuations in the previous cases. 
From this reason, we would like to name the structural quantum criticality. 

\subsubsection{Correlation between elastic anomaly and superconductivity}
\begin{figure}[t]
\begin{center}
\includegraphics[width=8.5cm]{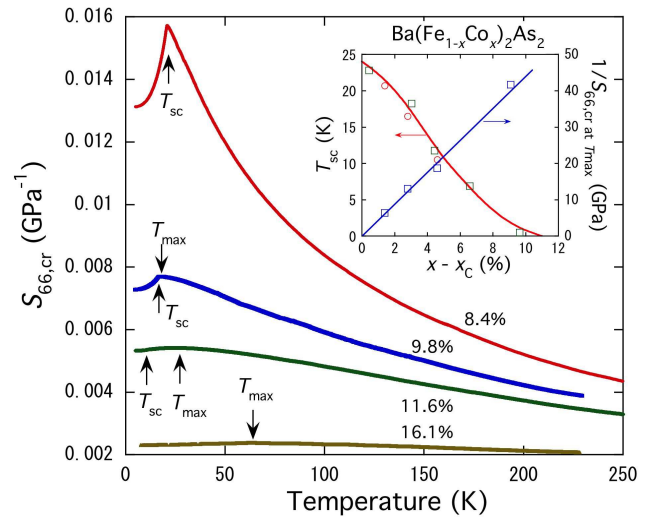}
\end{center}
\caption{(Color online) Temperature dependence of $S_{66,cr}$ for the overdoped samples.
 Inset is $T_{\rm sc}$ from this work and Ref. [25], and the inverse of the peak value of $S_{66,cr}$ as a function of the distance from the QCP concentration $x - x_{\rm C}$.
 The curve of $T_{\rm sc}$ is the guide for eyes.}
\label{S66}
\end{figure}

\begin{figure}[b]
\begin{center}
\includegraphics[width=8.5cm]{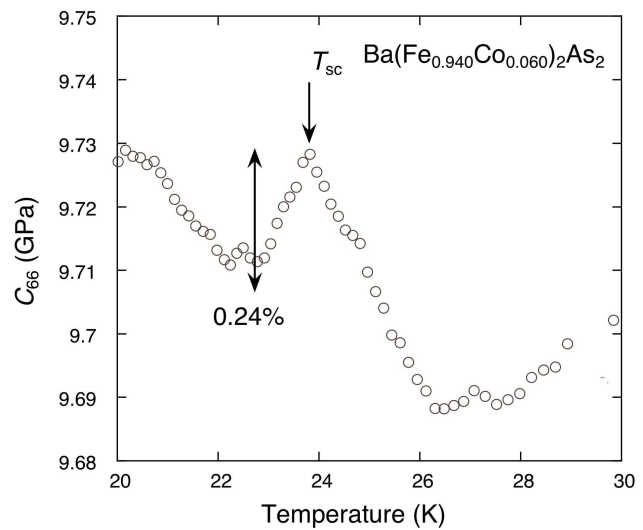}
\end{center}
\caption{(Color online) Temperature dependence of $C_{66}$ for the overdoped Ba(Fe$_{0.940}$Co$_{0.060}$)$_2$As$_2$ near $T_{\rm sc}$ in expanded scales}
\label{C66large}
\end{figure}

We now discuss the relationship between the elastic anomaly and superconductivity. 
As seen in the inset of Fig. \ref{S66}, the amount of 1/$S_{66,cr}$ is proportional to $x - x_{\rm C}$, where $x_{\rm C}$ is the QCP concentration of Co; $x_{\rm C}$ = 0.07 for this system.
Such behavior is well known in $\chi$ of the magnetic QCP. 
It is surprising that such well-known behavior holds in  this system with the respect of $S_{66}$ instead of $\chi$ for the magnetic system.
As shown in the same figure, $T_{\rm sc}$ decreases with increasing $x - x_{\rm C}$.
So we can recognize an apparent correlation between $T_{\rm sc}$ and 1/$S_{66,cr}$ that $T_{\rm sc}$ is a function of 1/$S_{66,cr}$
The explanation for this interesting fact is speculated to be as follows. 
As shown in Fig. \ref{C66large}, the underdoped sample exhibits a small anomaly at $T_{\rm sc}$, while a large upturn at $T_{\rm sc}$ is seen in the overdoped samples. 
Once the system enters the orthorhombic phase from the tetragonal phase, structural fluctuations are suppressed in the ordered phase. 
In the overdoped samples, however, structural fluctuations still survive even at $T_{\rm sc}$. 
The amount of the anomaly at $T_{\rm sc}$ correlates with the peak height of $S_{66,cr}$, which is the measure of structural fluctuation. 
The large anomaly at $T_{\rm sc}$ for the overdoped samples suggests strong coupling between the structural fluctuations and superconductivity. 
These facts suggest that the origin of $S_{66,cr}$ is deeply related to the emergence of superconductivity.

\section{Summary and concluding remarks}
The microscopic origin of $C_{66}$ softening has been discussed based on various mechanisms. 
Large elastic anomalies are often observed in the 4f electron system with a bilinear coupling between quadrupole operator $O$ and elastic strain $\varepsilon$ with the form of $O\varepsilon$. 
First, the effect of spin-nematic order has been considered, where the elastic anomaly above $T_{\rm N}$ is caused by the fluctuation of a pair of magnons at two sublattices \cite{cano,fernandes}. 
Spin-nematic order parameter can couple to the strain with a bilinear form. 
In this case, the nematic order deforms the crystal through the spin-orbit interaction. 

On the other hand, theories based on orbital fluctuation have been proposed.
Yanagi {\it et al}. considered Ferro-orbital order accompanying large elastic softening \cite{yanagi3}. 
Kontani {\it et al}. have discussed a two-orbiton process, which is initiated by antiferroquadrupolar fluctuation caused by interband nesting and consisting of a coupling between optical phonon and orbiton at the zone-boundary. 
They showed that two-orbiton process  brings about very large $C_{66}$ softening \cite{kontani2}. 

As regards the values of the electron-lattice coupling constants $\lambda$, the spin-nematic theory and the orbital theory give 17 meV/Fe and 0.2 eV/Fe, respectively \cite{fernandes,kontani2}.
Our experiment provides the value $\lambda$ $\sim$ 0.22-0.25 eV/Fe for the underdoped samples, and $d$ $\sim$ 0.2-0.28 eV/Fe for the overdoped samples.
These values of the coupling constant are considered to be very large, and are consistent with the orbital-based theory both from the localized and band pictures. 
Generally speaking, a coupling between the magnetic moment and the stain is mediated by a spin-orbit interaction. 
It is considered that such strong electron-lattice coupling is not caused by weak spin-orbital coupling in transition metals, but by the orbital nature of this system. 
We suppose that the strong electron-lattice coupling is a characteristic feature of multi-band systems, possessing orbital degrees of freedom.
Our results are considered to support the orbital-based theory. 

In this article, structural quantum critical behavior was reported for Ba(Fe$_{1-x}$Co$_x$)$_2$As$_2$.
The QCP behavior has been also reported by resistivity measurements and NMR of Ba(Fe$_{1-x}$Co$_x$)$_2$As$_2$ \cite{doiron,ning}. 
The QCP behavior observed in NMR and elastic measurements would be expected to be ascribed to the same origin. 
Elastic constant is no sensitive probe for magnetism, but sensitive to orbital (quadrupole), contrary to NMR. 
Our measurement is the observation from the side of orbital (quadrupole). 

Further discussions will be necessary on the role of the spin and orbital in the mechanism of superconductivity. 
However, this system was found to be well characterized by a strong electron-lattice coupling. 
Therefore, irrespective of whether the spin or orbital mediate superconductivity, the results of the present study suggest that the structural fluctuation must be actively incorporated into their role, and that the orbital degrees of Fe-3d must also be taken into consideration in order to understand the full picture of superconductivity in iron-based compounds.

\section*{Acknowledgments}
The authors wish to thank H. Fukuyama, H. Hosono, Y. {\~ O}no, H. Kontani, Y. Yanagi, J. Schmalian and L{\" u}thi for their valuable discussions; and M. Nakamura and T. Kowata for their assistance in the experiments.
This work was supported by a Grant-in-Aid for Scientific Research on Innovative Areas, gHeavy Electronsh (No. 20102007), of The Ministry of Education, Culture, Sports, Science and Technology, Japan, and the Transformative Research-project on Iron Pnictides of the Japan Science and Technology Agency.


\begin{thebibliography}{9}
\bibitem{jonston} D. C. Johnston, Adv. Phys. 59 (2010) 803.
\bibitem{rotter} M. Rotter, M. Tegel, D. Johrendt, I. Schellenberg, W. Hermes, and R P{\" o}ttgen, Phys. Rev. B 78 (2008) 020503(R). 
\bibitem{huang}Q. Huang, Y. Qiu, Wei Bao, M.A. Green, J.W. Lynn, Y.C. Gasparovic, T. Wu, G. Wu and X. H. Chen, Phys. Rev. Lett. 101 (2008) 257003. 
\bibitem{sefat} A. S. Sefat, R. Jin, M. A. McGuire, B. C. Sales, D. J. Singh and D. Mandrus, Phys. Rev. Lett. 101 (2008) 117004.
\bibitem{rotter2} M. Rotter, M. Tegel and D. Johrendt, Phys. Rev. Lett. 101 (2008) 107006.
\bibitem{mathur} N. D. Mathur, F. M. Grosche, S. R. Julian, I. R. Walker, D. M. Freye, R. K. W. Haselwimmer, and G. G. Lonzarich, Nature 394 (1998) 39.
\bibitem{saxena} S. S. Saxena, P. Agarwal, K. Ahilan, F. M. Grosche, R. K. W. Hasselwimmer, M. J. Steiner, E. Pugh, I. R. Walker, S. R. Julian, P. Monthoux, G. G. Lonzarich, A. Huxley, I. Sheikin, D. Braithwaite and J. Flouquet, Nature 406 (2000) 587.
\bibitem{fang} C. Fang, H. Yao, W. F. Tsai, J-P. Hu, and St. A. Kivelson, Phys. Rev. B 77 (2008) 224509.
\bibitem{krueger} F. Kr{\" u}ger, S. Kumar, J. Zaanen, and J. van den Brink, Phys. Rev. B 79 (2009) 054504.
\bibitem{mazin} I. I. Mazin, D. J. Singh, M. D. Johannes, and M. H. Du, Phys. Rev. Lett. 101 (2008) 057003. 
\bibitem{kuroki} K. Kuroki, S. Onari, R. Arita, H. Usui, Y. Tanaka, H. Kontani, and H. Aoki, Phys. Rev. Lett. 101 (2008) 087004. 
\bibitem{kontani1} H. Kontani and S. Onari, Phys. Rev. Lett. 104 (2010) 157001. 
\bibitem{yanagi1} Y. Yanagi, Y. Yamakawa, and Y. {\~ O}no, Phys. Rev. B 81 (2010) 054518.
\bibitem{fernandes} R. M. Fernandes, L. H. Van Bebber, S. Bhattacharya, P. Chandra, V. Keppens, D. Mandrus, M. A. McGuire, B. C. Sales, A. S. Sefat, and J. Schmalian, Phys. Rev. Lett. 105 (2010) 157003. 
\bibitem{goto} T. Goto, R. Kurihara, K. Araki, K. Mitsumoto, M. Akatsu, Y. Nemoto, S. Tatematsu, and M. Sato, J. Phys. Soc. Jpn. 80 (2011) 073702.
\bibitem{luethi1} B. L{\" u}thi, Physical Acoustics in the Solid State, Springer-Verlag p. 9 (2004).
\bibitem{yoshizawa_news} M. Yoshizawa, J. Phys. Soc. Jpn. Online-News and Comments [July 14, 2011]. 
\bibitem{testerdi} L. R. Testerdi, and T. B. Bateman, Phys. Rev. 154 (1967) 402.
\bibitem{suzuki} T. Suzuki, H. Goshima, S. Sakita, T. Fujita, M. Hedo, Y. Inada1, E. Yamamoto, Y. Haga and Y. {\^ O}nuki, J. Phys. Soc. Jpn. 65 (1996) 2753.
\bibitem{thalmeier} P. Thalmeier and B. L{\" u}thi, in Handbook on the Physics and Chemistry of Rare Earths Vol. 14, ed. K. A. Gschneidner Jr. and L. Eyring, Amsterdam: Elsevier (1991) p. 245.
\bibitem{yoshizawa_bp} M. Yoshizawa, I. Shirotani, and T. Fujimura, J. Phys. Soc. Jpn. 55 (1986) 1196.
\bibitem{luethi159} B. L{\" u}thi, Physical Acoustics in the Solid State, Springer-Verlag  (2004) p. 159.
\bibitem{rullier} F. Rullier-Albenque et al., Phys. Rev. Lett. 103 (2009) 057001.
\bibitem{nakajima} M. Nakajima, S. Ishida, K. Kihou, Y. Tomioka, T. Ito, Y. Yoshida, C. H. Lee, H. Kito, A. Iyo, H. Eisaki, K. M. Kojima, and S. Uchida, Phys. Rev. B 81 (2010) 104528.
\bibitem{canfield} P. C. Canfield, S. L. Budfko, Ni Ni, J. Q. Yan, and A. Kracher, Phys. Rev. B 80 (2009) 060501(R).
\bibitem{laplace} Y. Laplace, J. Bobroff, F. Rullier-Albenque, D. Colson, and A. Forget, Phys. Rev. B 80 (2009) 140501(R).
\bibitem{doiron}N. Doiron-Leyraud, P. Auban-Senzier, S. R. de Cotret, Cl. Bourbonnais, D. J{\' e}rome, K. Bechgaard, and L. Taillefer, Phys. Rev. B 80 (2009) 214531. 
\bibitem{kasahara} S. Kasahara, T. Shibauchi, K. Hashimoto, K. Ikada, S. Tonegawa, R. Okazaki, H. Shishido, H. Ikeda, H. Takeya, K. Hirata, T. Terashima, and Y. Matsuda, Phys. Rev. B 81 (2010) 184519.
\bibitem{onuki}Y. {\^ O}nuki, R. Settai, K. Sugiyama, T. Takeuchi, T. C. Kobayashi, Y. Haga, and E. Yamamoto, J. Phys. Soc. Jpn. 73 (2004) 769.
\bibitem{nakai} Y. Nakai, T. Iye, S. Kitagawa, K. Ishida, H. Ikeda, S. Kasahara, H. Shishido, T. Shibauchi, Y. Matsuda, and T. Terashima, Phys. Rev. Lett. 105 (2010) 107003.
\bibitem{cano} A. Cano, M. Civelli, I. Eremin, and I. Paul, Phys. Rev. B 82 (2010) 020408(R). 
\bibitem{yanagi3} Y. Yanagi, Y. Yamakawa, N. Adachi, and Y. {\~ O}no, J. Phys. Soc. Jpn. 79 (2010) 123707. 
\bibitem{kontani2} H. Kontani, T. Saito, and S. Onari, Phys. Rev. B 84 (2011) 024528.
\bibitem{ning}F. L. Ning, K. Ahilan, T. Imai, A. S. Sefat, M. A. McGuire, B. C. Sales, D. Mandrus, P. Cheng, B. Shen, and H.-H Wen, Phys. Rev. Lett. 104 (2010) 037001.
\end{thebibliography}
\end{document}